\documentclass[twocolumn,showpacs,prl]{revtex4-1}
\usepackage{graphicx}
\usepackage{rotating}
\usepackage{color}
\usepackage{amssymb}
\newcommand{\chib}{\mbox{\raisebox{0.3ex}{$\chi$}}_{b}}
\newcommand{\bmprl}[1]{\mbox{\boldmath $#1$}}
\begin{document}
\title{Comment on ``Observation of a New
\ensuremath{\raise.4ex\hbox{\boldmath $\chi$}_{{\boldmath b}}}
State in Radiative Transitions to \bmprl{\Upsilon}(1S)
and \bmprl{\Upsilon}(2S) at ATLAS''}
\author{
Eef~van~Beveren$^{\; 1}$ and George~Rupp$^{\; 2}$}
\affiliation{
$^{1}$Centro de F\'{\i}sica Computacional,
Departamento de F\'{\i}sica,
Universidade de Coimbra, P-3004-516 Coimbra, Portugal\\
$^{2}$Centro de F\'{\i}sica das Interac\c{c}\~{o}es Fundamentais,
Instituto Superior T\'{e}cnico, Technical University of Lisbon,
P-1049-001 Lisboa Codex, Portugal
}
\pacs{
14.40.Pq, 
11.80.Gw, 
12.40.Yx, 
13.25.Gv 
}

\maketitle

We comment on the recent observation
of $\chib$ states
by the ATLAS Collaboration \cite{PRL108p152001} and contest
their peremptory interpretation of the highest
of the three observed structures as the $\chib$(3P) system. Moreover, we
do not agree that from their data the mass barycenter
of the $\chib$(3P) system can be determined.

\begin{figure}[htbp]
\begin{center}
\includegraphics[width=240pt, angle=0]{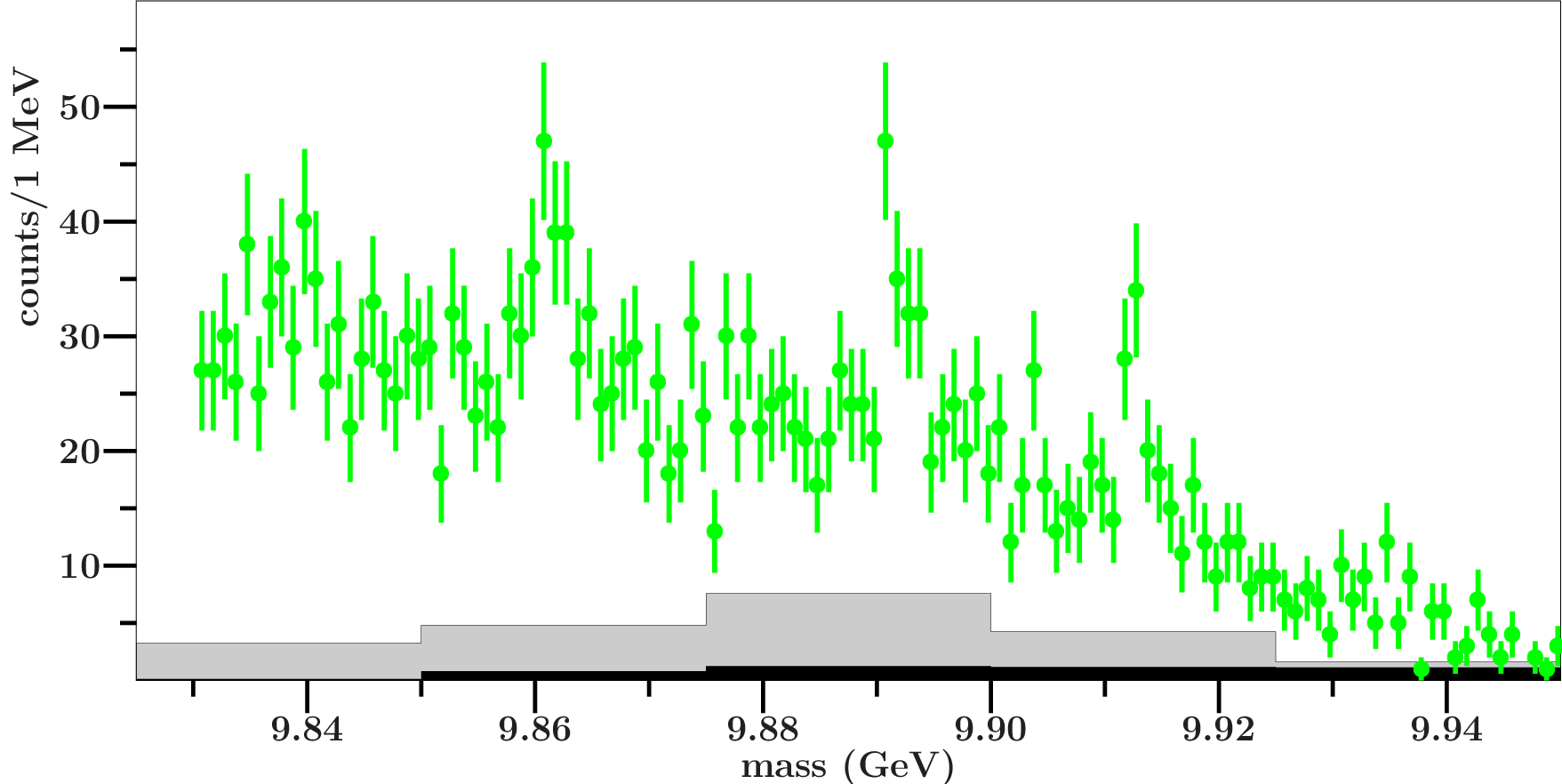}
\end{center}
\caption[]{Photon counts per 1 MeV,
assuming transitions $\chib(1P)\to\Upsilon (1S)\gamma$
for converted (gray) and unconverted photons (black)
as shown by the ATLAS Collaboration \cite{PRL108p152001},
compared to the photon distribution measured by
the ARGUS Collaboration \cite{PLB160p331}
(dots and error bars).}
\label{chib1Pstates}
\end{figure}

In the first place, it should be noticed that the ATLAS data for
the $\chib$(P) systems have a very poor energy resolution,
inhibiting an analysis of more detailed structures.
Figure~\ref{chib1Pstates} compares the presently available ATLAS data
for the $\chib$(1P) system with photon data obtained more than 25 years
ago \cite{PLB160p331}.
Furthermore, the ATLAS Collaboration justifies
their interpretation of the data by referring to theoretical work
\cite{PRD36p3401,EPJC4p107} that predicts the $\chib(3P)$ system
near the highest enhancement in their data \cite{PRL108p152001}.

Now, over many years we have been advocating that me\-so\-nic
resonances are not pure (``quenched'') $q\bar{q}$ states, but
can have large meson-meson components, too. Thus, bottomonium
systems also contain contributions of open-bottom meson pairs.
An adequate approach to study such effects is a coupled-channel
$T$-matrix formalism \cite{PRD84p094020}, even below the lowest
open-bottom threshold ($B\bar{B}$). When applied to the referred
theoretical predictions \cite{PRD36p3401,EPJC4p107}, the results
will change substantially \cite{PRD27p1527,ARXIV11050855}.

Moreover, the unconverted ATLAS photon data
\cite{PRL108p152001} show more structure in the energy region
above the $B\bar{B}$ threshold than accounted for in their fit to the
data. Indeed, photonic decays will certainly be hindered by strong
open-bottom decays above the $B\bar{B}$ threshold.
Nevertheless, it is exactly in that region where we expect
a triplet $(0^{++},1^{++},2^{++})$ of $\chib$ states.

\begin{figure}[htbp]
\begin{center}
\includegraphics[width=220pt, angle=0]{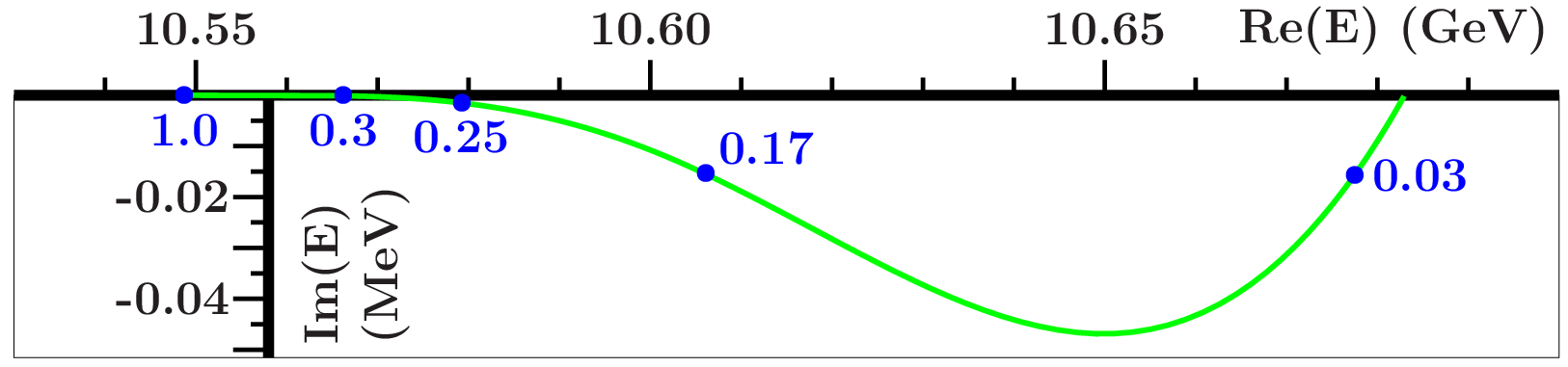}
\end{center}
\caption[]{Resonance-pole movement for lowest
$3^{\, 3\!}P_{2}\,$-$\,2^{\, 3\!}F_{2}$ state,
as a function of gradual unquenching.
Numbers along curve indicate degree of unquenching.}
\label{2Fpole}
\end{figure}

Using the model and parameters of Ref.~\cite{PRD27p1527},
we find \em four \em \/quenched $\chib$ candidates
at 10.683 GeV, viz.\ the three $\chib\left( 3^{\, 3\!}P_{0,1,2}\right)$
states, but also the $\chib\left( 2^{\, 3\!}F_{2}\right)$ $2^{++}$, which
will mix with $\chib\left( 3^{\, 3\!}P_{2}\right)$ upon unquenching
\cite{PRD84p094020}. One of the resulting
$3^{\, 3\!}P_{2}\,$-$\,2^{\, 3\!}F_{2}$ combinations
comes out around 10.62 GeV, just like the unquenched
$3^{\, 3\!}P_{0}$ and $3^{\, 3\!}P_{1}$ states,
whereas the other $2^{++}$ mixture ends up as a bound state,
below the $B\bar{B}$ threshold, at 10.548 GeV (see Fig.~\ref{2Fpole}).

We thus conclude that the ATLAS Collaboration \cite{PRL108p152001}
may very well have observed a single
$3^{\, 3\!}P_{2}\,$-$\,2^{\, 3\!}F_{2}$ state.

This important issue can be settled with a much higher energy
resolution, showing whether there is a triplet of $J^{++}$ states
around 10.54 GeV or just a singlet.

\newcommand{\pubprt}[4]{{#1 {\bf #2}, #3 (#4)}}
\newcommand{\ertbid}[4]{[Erratum-ibid.~{#1 {\bf #2}, #3 (#4)}]}
\def\EPJC{Eur.\ Phys.\ J.\ C}
\def\PLB{Phys.\ Lett.\ B}
\def\PRD{Phys.\ Rev.\ D}
\def\PRL{Phys.\ Rev.\ Lett.}

\end{document}